\documentclass{iau}

\usepackage{amsmath}
\usepackage{graphicx}
\usepackage{multirow}

\newcommand\sbullet[1][.5]{\mathbin{\vcenter{\hbox{\scalebox{#1}{$\bullet$}}}}}

\begin{document}

\lefttitle{M. Wielgus \& A. Yfantis}
\righttitle{Dissecting Sgr~A* variability in the orbiting hotspot framework}

\jnlPage{1}{7}
\jnlDoiYr{2021}
\doival{10.1017/xxxxx}

\aopheadtitle{Proceedings IAU Symposium}
\editors{M. Zaja\v{c}ek,  T. Je\v{r}\'{a}bkov\'{a}, V. Karas, R. Schödel \&  P. Sukov\'{a}, eds.}

\title{Dissecting the variability of Sagittarius A* \\ in the orbiting hotspot framework}

\author{Maciek Wielgus and Aristomenis I. Yfantis}
\affiliation{Instituto de Astrofísica de Andalucía-CSIC, Glorieta de la Astronomía s/n, E-18008 Granada, Spain \email{maciek@wielgus.info }\email{; ayfantis@iaa.csic.es}}

\begin{abstract}

Sagittarius~A* (Sgr~A*) exhibits intermittent flaring behavior, characterized by strong enhancements in high-energy emission (infrared and X-ray). These events have been interpreted as related to the coherent orbital motion of localized regions of energized plasma -- hotspots. While this framework has been quite fruitful in constraining properties of Sgr~A*, it is subject to important challenges and limitations. In this brief review we discuss the recent observational and theoretical developments related to this line of research, commenting on the concerns regarding the hotspot model validity and the future perspectives.

\end{abstract}

\begin{keywords}
Sagittarius~A*, polarimetry, Galactic Center, Astrophysical flares
\end{keywords}

\maketitle

\section{Introduction}

Our closest supermassive black hole, Sagittarius~A* (Sgr~A*), located in the center of the Milky Way, constitutes a unique laboratory for probing relativistic accretion physics and strong-field gravity. With a mass of $M_{\sbullet[1.0]} = 4.3 \times 10^6\,M_{\odot}$ \textcolor{red}{} \citep{Gravity2022} the associated dynamical timescale (Keplerian period on the innermost stable circular orbit) is only $\sim 0.5\,$h. It has been demonstrated that millimeter-wavelength and near-infrared (NIR) emission originates from the innermost region of the system, within a few gravitational radii ($r_g = G M_{\sbullet[1.0]} / c^2 $) from the central object \citep{gravity:2018,eht:2022_paperI}. Hence, one can reasonably hope to find imprints of orbital dynamics in the observed mm and NIR light curves. 

A turbulent system is not strictly stationary and a transient bright inhomogeneity can briefly dominate its emission \citep{Sunyaev1972}. Sgr~A* exhibits irregular flaring activity, manifesting clearly in X-ray and NIR observations through an increase of the instantaneous flux density by 1–2 orders of magnitude \citep{Baganoff2001,Genzel2003}. These flaring episodes are likely related to dissipative events energizing plasma locally and thus enabling the formation of such bright localized features. Subsequently, the observed signatures of such an energized \emph{hotspot}, executing orbital motion while embedded in the background accretion flow, could be used as a probe of the system properties, including its geometry and dynamics \citep{Cunningham1972,Bao1992,Brod-Loeb2006, Trippe2007, Dovciak2008, Hamaus2009}. 

To date, the most convincing argument for the hotspot scenario is given by the NIR observations of \cite{gravity:2018}, demonstrating astrometric rotation of the brightness centroid during the flaring event and simultaneous rotation of the linear polarization (electric vector position angle; EVPA) on a plane of Q, U Stokes parameters (QU-loops). A coherent polarized feature at a compact orbital radius $r_{\rm orb} \sim 9\, r_g$ moving clockwise through an axisymmetric poloidal magnetic field and observed at low inclination offers a simple and alluring interpretation of these observations \citep{gravityMichi,gravityAlejandra}.

But do such localized orbiters really form around Sgr~A* and is their behavior truly coherent enough to teach us something about their environment? Or, alternatively, are we perhaps reading too much into sparse stochastic data, subject to selection bias and applying our oversimplified models to a complex turbulent system? In this short contribution we attempt to address these issues critically, emphasizing recent developments and future perspectives.

\section{Recent developments (observations, inference, and theory)}

\begin{figure}[h!]
    \centering
    \includegraphics[width=0.3\linewidth,trim={0.6cm 0.cm 0 0.cm},clip]{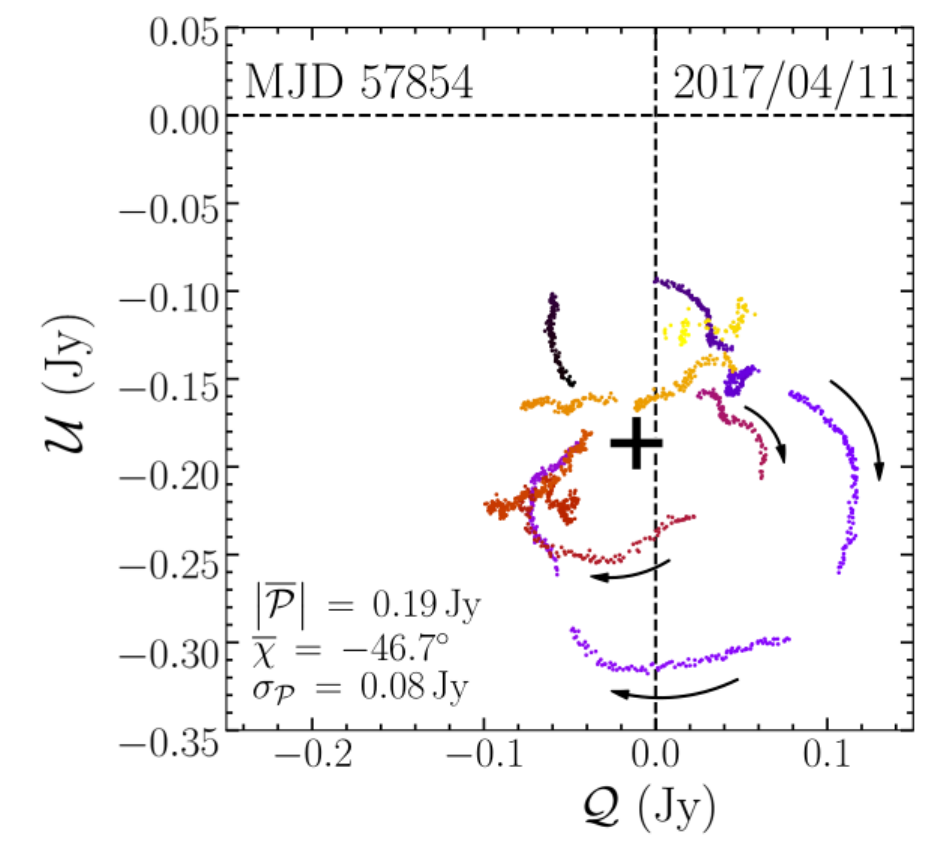}
    \includegraphics[width=0.304\linewidth,trim={1.8cm 2.1cm 2.56cm 0.9cm},clip]{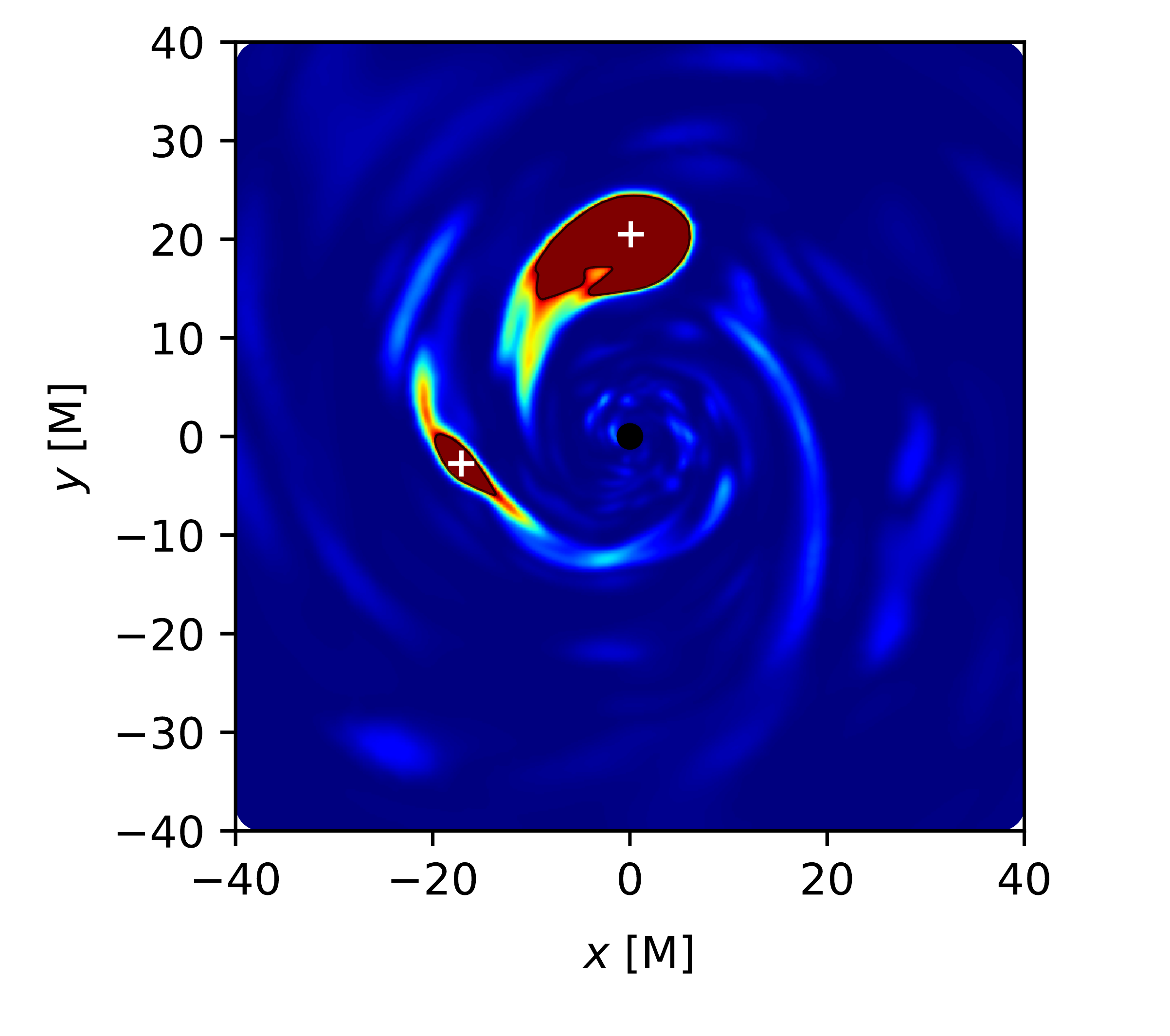}
    \includegraphics[width=0.302\linewidth,trim={0.cm 0.cm 3.22cm 0.cm},clip]{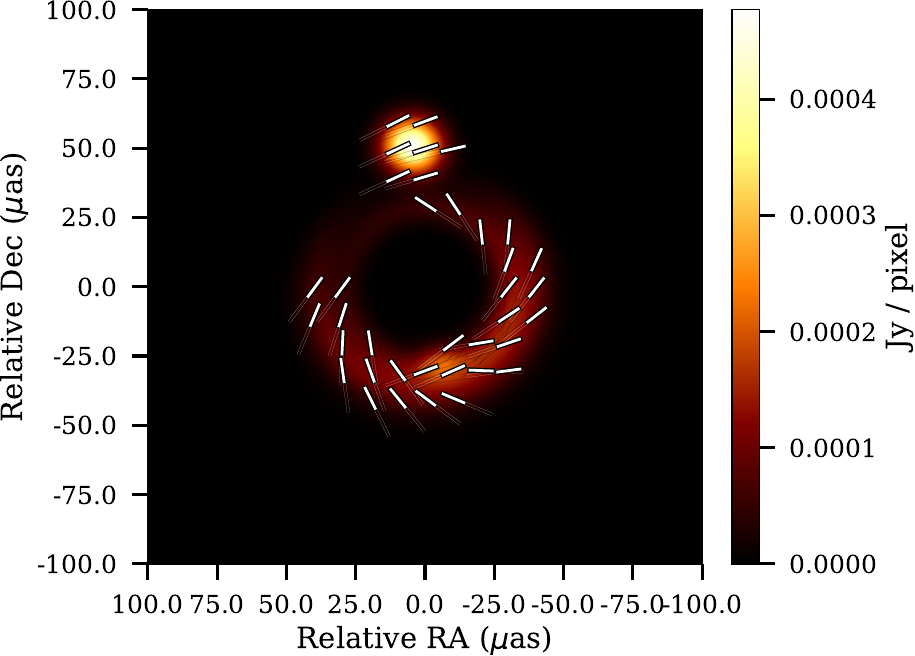}

    \caption{Left: ALMA observation of QU-loops \citep{W22}. Middle: equatorial slice from a GRMHD simulation, color-coded for $B_z^2/P_{\text{gas}}$ \citep{Porth2021_fluxer}, showing the formation of orbiting flux tubes -- hotspot candidates. Right: ray-traced image of a geometric hotspot model with an accretion disk background \citep{yfantis24a}. Ticks indicate the direction of the linear polarization. }
    
    \label{fig:hotspot}
\end{figure}

An exciting new avenue for understanding variability in Sgr~A* emerged with the identification of polarization signatures of an orbiting hotspot in the Atacama Large Millimeter/submillimeter Array (ALMA) observations at $230$ GHz \citep{W22}, associated with an X-ray flare observed by Chandra \citep{eht:2022_paperII,Wielgus2022_LC}, see the left panel of Fig.~\ref{fig:hotspot}. Compared to the NIR observations from GRAVITY, these measurements provided a substantially higher signal-to-noise ratio, usable absolute Stokes Q, U values (instead of fractional Q/I, U/I), and much denser temporal sampling. In parallel, \citet{Tiede2020}, \citet{Gelles:2021}, and \citet{Vos:2022} explored the detailed predictions of the hotspot model more systematically, leading to an interpretation of the ALMA linear polarization light curves based on the analysis of \citet{Vos:2022} given in \citet{W22}. Broad consistency with the GRAVITY observations was established, particularly regarding low orbital inclination, direction of motion, and presence of a strong poloidal magnetic field. 

To systematically explore consistency between data and geometric hotspot models, such as the one ray-traced in the right panel of Fig.~\ref{fig:hotspot}, \citet{yfantis24a} developed a Bayesian fitting algorithm \texttt{bipole}, extending the capabilities of the general relativistic ray-tracing code \texttt{ipole} \citep{Monika2018}. Applied to ALMA observations, the statistical analysis confirmed the overall consistency with \citet{W22}. Apart from that, a~strong preference for the sub-Keplerian orbital velocity was indicated for the first time, with $\Omega \sim 0.7 \, \Omega_{\rm Kep}$. No strong evidence of the impact of cooling on dynamical timescales was found, putting mild constraints on the magnetic field strength. The orbital radius was estimated to be $r_{\rm orb} \sim 11\, r_g$, somewhat larger than that inferred from the NIR observations. \citet{Aviad2024} took a different approach to ALMA data, with weak assumptions on the emitting region morphology, using a deep learning algorithm to recover a coherent orbiter with similar parameters. 
 
\citet{gravity2023} presented an ensemble average of eight NIR flaring events, six with polarimetric and four with astrometric data (and with two datasets containing both types of data), from observations between 2018 and 2022. The dataset should increase the inferential power of the hotspot models, assuming, of course, that there exists a well-defined average NIR hotspot event, corresponding to some characteristic orbital parameters. Here \texttt{bipole} enabled simultaneous fitting of polarimetric and astrometric observations \citep{yfantis24b}. The analysis disfavored sub-Keplerian orbital velocities, resulting in a tension with the constraints from ALMA data. Overall, the mm wavelength event of \citet{W22} appears to involve a hotspot orbiting at a larger radius, with a longer period and a more sub-Keplerian velocity than the NIR events. Another tension is related to the mismatch of the estimated position angle of the screen-projected orbital angular momentum vector, see \citet{gravity2023,yfantis24a,yfantis24b,janssen25c_pa135}. 
There is, however, an additional complication for the ALMA data analysis related to the EVPA rotation originating in a mixed internal / external Faraday screen \citep{Wielgus2023}.

On the theoretical front the relation between the magnetic flux eruptions in magnetically arrested disks (MADs; \citealt{Narayan2003}) and flaring episodes has been extensively explored \citep[e.g.,][]{dexter20,Porth2021_fluxer,antonop025_spiral}. In this scenario magnetic field accumulates near the horizon, reducing the accretion rate and irregularly reconfiguring in flux eruption events in the presence of plasma instabilities and vigorous reconnection \citep{Ripperda_2022}. Plasmoids and flux tubes are formed in such events and constitute good candidates for a theoretical interpretation of localized hotspots that could be advected with the flow \citep[e.g.,][]{Ripperda_2020, Miki2022}. An example of such localized features forming in a general relativistic magnetohydrodynamic (GRMHD) simulation is shown in the middle panel of Fig.~\ref{fig:hotspot} \citep{Porth2021_fluxer}. Overall, while the detailed and quantitative physics of this MAD cycle requires further investigation, the connection to flaring in Sgr~A* is quite convincing, as the process is ubiquitous in GRMHD simulations, characteristic of the compact environment, and generates repetitive irregular events during which magnetic energy is dissipated, locally heating and accelerating plasma. 

Lastly, some recent theoretical works focused on the prospects for observational tests of strong gravity within the orbiting hotspot paradigm. The observable signatures are sensitive to the spacetime curvature \citep[e.g.,][]{Brod-Loeb2006, Vincent2023} and hence they may enable probing black hole spin under the Kerr metric assumption \citep{Vos:2022,yfantis24a}. Alternatively, deviations from Kerr also manifest in the observables \citep{Shahzadi2022,Aimar2026}. The perspective of detecting the second-order image of the hotspot, formed by photons looping around the central ultra-compact object, offers a~particularly attractive possibility of probing the geometry through its temporal and angular delays with respect to the direct image \citep{Prashant2024,Y26}.

\section{The hotspot model criticism and challenges}
\label{sec:criticism}

Despite the existing observations, the connection between energetic flares and coherent orbital motion is not unanimously accepted. The concerns are related to a possible selection bias in the observed data sets as well as to employing rather simplistic geometric models that may impose the hotspot interpretation. However, recent theoretical works strengthened this association. While \citet{dexter20} demonstrated polarimetric and astrometric looping in simulated NIR light curves during flux eruption events, \citet{Najafi2024} recreated QU-loops similar to those observed by ALMA in their GRMHD simulations. Identification of a single bright and coherent orbiter in the ray-traced simulations is rarely straightforward though and hence more work involving modeling of hotspots in GRMHD synthetic data is needed. \citet{Ricarte2025} showed a statistical analysis of QU-looping in polarimetric ALMA data beyond the flaring state, finding a degree of consistency between observations and numerical GRMHD models, particularly those with clockwise accretion observed at low inclination.

In the flux eruption interpretation, there is nothing to guarantee the formation of a single hotspot dominating the observed variable flux density. The presence of several energized orbiting features would obscure both astrometric and polarimetric observables. This may justify the selective choice of the considered data sets. As an example, in April 2018 ALMA once again observed Sgr~A* in the aftermath of an X-ray flare \citep{Ezequiel2026}, but unlike in the observations of \citet{W22} no clear QU-loop pattern was identified this time, although some statistical differences from quiescent observing days have been reported. 

While simple geometric models commonly consider a circular equatorial orbit of a hotspot, in GRMHD simulations plasmoids and flux tubes are often moving out of the system entrained in the outflowing jet, or falling into the black hole with the accretion disk \citep[e.g.,][]{Ripperda_2020,Ripperda_2022}. The simple models appear to represent ALMA and GRAVITY data sufficiently well in a statistical sense \citep{yfantis24a,yfantis24b}, nonetheless this may just indicate the lack of inferential power in these datasets and it is important to explore other dynamics suggested by numerical simulations. Some off-equatorial outflowing models for GRAVITY astrometric data were discussed by \citet{ball:2021,aimar_2023_plasmoid,Antonopoulou24}. In particular, the properties of these models may explain the preference for super-Keplerian motion in the fits assuming an equatorial model \citep{yfantis24b}. Recently \citet{Chen2025_inspiral,2026ruales_inspiral} calculated polarimetric signatures of inspiralling hotspots. These models should be tested against the ALMA observations. It is also entirely possible that some events lead to the ejection of a feature while others lead to an infall, which could explain tentative inconsistencies between events observed in NIR and the ALMA event. Another possible limitation is that geometric models rarely account for the evolution of the hotspot state in its comoving frame (differential flow shearing, cooling), with notable exceptions such as \citet{Tiede2020,yfantis24a}. Hence, the important underlying assumption is that the timescale for this internal evolution is longer than the dynamical timescale.

Apart from the criticism of the geometric models, there are also concerns related to treating GRMHD simulations as a realistic model of Sgr~A*, with which we could justify and understand the application of the simple geometric models. Some of these concerns pertain to the commonly discussed general limitations of the GRMHD approach, such as using the fluid approach to describe collisionless plasma, particularly in the magnetically-dominated low-density regions, or employing unrealistic initial conditions. In the context of flaring behavior specifically, simplified prescription-based modeling of electron heating constitutes a relevant weak point. The electron energy distribution is also assumed, and in most cases simply modeled with the Maxwell–J\"{u}ttner distribution \citep{eht:2022_paperV, eht:2024b} thus lacking the high-energy tail for the energized electrons in the flaring state. While in a hotspot model this has little impact on the astrometric observables, it affects Q, U observables through the spectral index of the emission. The effect is more problematic at higher energies than at mm wavelengths. Furthermore, accurately accounting for magnetic reconnection may require using an extreme numerical resolution \citep{Ripperda_2022} and possibly relaxing the ideal (infinite conductivity) MHD approximation altogether. More broadly, these limitations of the fluid MHD approach motivate the use of computationally expensive kinetic simulations that do not enforce local charge neutrality. In a related recent study, \citet{Tlemissov2026} demonstrated the impact of an unscreened charge on the polarimetric observables from a geometric model of an orbiting hotspot.

Finally, while the accretion flow in Sgr~A* is very radiatively inefficient and generally it is not relevant to include dynamical coupling between plasma and radiation in the simulations of Sgr~A*, radiative cooling almost certainly plays a role for the energized flow features -- tentative hotspots. In particular, radiative losses following the flare likely inform the temporal evolution of the energy spectrum thus impacting multiwavelength characteristics of a flaring event. \citet{yfantis24a} argued that the cooling timescale may be sufficiently long at 230\,GHz, but it is definitely important for the NIR data. In practice, using fractional polarization components (Q/I, U/I) partly mitigates these effects.

\section{Future perspectives}

The ongoing upgrades to the Event Horizon Telescope (EHT) array should enable resolving mm wavelength structure of the source both spatially and in time, including polarization \citep{ngeth23}. Thus, presence of a localized, dominant orbiting component in the high-energy flare aftermath and its relation to the background accretion flow could be tested directly and robustly. Furthermore, precise tracking of the orbital dynamics and internal evolution (cooling, expansion, impact of instabilities, smearing in a differential flow etc.) of a putative hotspot in such a ``black hole video" should become possible \citep{Emami2023}. If such observations enable the primary and secondary images to be spatially isolated, a new sensitive probe into the spacetime geometry will be unlocked \citep{Y26}.

Additional observational progress is expected to result from the sheer accumulation of high-quality data sets across the frequency spectrum to improve the statistical models of the source behavior, dominated by the stochastic red noise component \citep{Wielgus2022_LC,Ezequiel2026}. Simultaneous multiwavelength observations, notably between infrared and mm, have a very important role to play especially in constraining the physics of the energy dissipation and plasma cooling as well as characterizing the Faraday screen of Sgr~A*. Altogether, the next decade of Sgr~A* observations should be most helpful, if not decisive, in clarifying the nature of flares and the validity of their interpretation in the hotspot framework. 

\section*{Acknowledgments}
\noindent This research benefited from the financial support from the Severo Ochoa grant CEX2021-001131-S funded by MCIN/AEI/ 10.13039/501100011033. MW is supported by a Ramón y Cajal grant RYC2023-042988-I from the Spanish Ministry of Science and Innovation. AIY acknowledges funding from the European Union’s Horizon Europe research and innovation program under grant agreement No. 101093934 (RADIOBLOCKS).

\bibliographystyle{iaulike}
\bibliography{Sample} 

@ARTICLE{janssen25c_pa135,
       author = {{Janssen}, M. and {Chan}, C.-k. and {Davelaar}, J. and {Wielgus}, M.},
        title = "{Deep learning inference with the Event Horizon Telescope: III. ZINGULARITY results from the 2017 observations and predictions for future array expansions}",
      journal = {A\&A},
         year = 2025,
        month = jun,
       volume = {698},
          eid = {A62},
        pages = {A62},
          doi = {10.1051/0004-6361/202553786},
archivePrefix = {arXiv},
       eprint = {2506.13877},
 primaryClass = {astro-ph.HE},
       adsurl = {https://ui.adsabs.harvard.edu/abs/2025A&A...698A..62J}
}

@ARTICLE{2026ruales_inspiral,
       author = {{Ruales}, Pablo and {Gates}, Delilah E.~A. and {C{\'a}rdenas-Avenda{\~n}o}, Alejandro},
        title = "{Polarization signatures of inspiraling hotspots around Kerr black holes}",
      journal = {PRD},
         year = 2026,
        month = may,
       volume = {113},
       number = {10},
          eid = {103030},
        pages = {103030},
          doi = {10.1103/65fp-f288},
archivePrefix = {arXiv},
       eprint = {2602.09102},
 primaryClass = {astro-ph.HE},
       adsurl = {https://ui.adsabs.harvard.edu/abs/2026PhRvD.113j3030R}
}

@ARTICLE{Chen2025_inspiral,
       author = {{Chen}, Bin and {Hou}, Yehui and {Song}, Yu and {Zhang}, Zhenyu},
        title = "{Polarization patterns of the hot spots plunging into a Kerr black hole}",
      journal = {PRD},
         year = 2025,
        month = apr,
       volume = {111},
       number = {8},
          eid = {083045},
        pages = {083045},
          doi = {10.1103/PhysRevD.111.083045},
archivePrefix = {arXiv},
       eprint = {2407.14897},
 primaryClass = {astro-ph.HE},
       adsurl = {https://ui.adsabs.harvard.edu/abs/2025PhRvD.111h3045C}
}

@ARTICLE{antonop025_spiral,
       author = {{Antonopoulou}, Eleni and {Loules}, Argyrios and {Nathanail}, Antonios},
        title = "{Magnetically arrested disk flux eruption events to describe SgrA* flares}",
      journal = {A\&A},
         year = 2025,
        month = apr,
       volume = {696},
          eid = {A10},
        pages = {A10},
          doi = {10.1051/0004-6361/202453456},
archivePrefix = {arXiv},
       eprint = {2501.07521},
 primaryClass = {astro-ph.HE},
       adsurl = {https://ui.adsabs.harvard.edu/abs/2025A&A...696A..10A}
}

@Article{ngeth23,
AUTHOR = {{Johnson}, M. D. and {Akiyama}, K. and {Bouman}, K. and {others}},
TITLE = {Key Science Goals for the Next-Generation Event Horizon Telescope},
JOURNAL = {Galaxies},
VOLUME = {11},
YEAR = {2023},
NUMBER = {3},
ARTICLE-NUMBER = {61},
URL = {https://www.mdpi.com/2075-4434/11/3/61},
ISSN = {2075-4434},
DOI = {10.3390/galaxies11030061}
}

@ARTICLE{Y26,
       author = {{Yfantis}, A.~I. and {Palumbo}, D.~C.~M. and {Mo{\'s}cibrodzka}, M.},
        title = "{Lensing of hot spots in Kerr space-time: An empirical relation for black hole spin estimation}",
      journal = {A\&A},
         year = 2026,
        month = feb,
       volume = {707},
          eid = {A35},
        pages = {A35},
          doi = {10.1051/0004-6361/202555277},
archivePrefix = {arXiv},
       eprint = {2504.16218},
 primaryClass = {astro-ph.HE},
       adsurl = {https://ui.adsabs.harvard.edu/abs/2026A&A...707A..35Y}
}

@ARTICLE{aimar_2023_plasmoid,
       author = {{Aimar}, N. and {Dmytriiev}, A. and {Vincent}, F.~H. and {El Mellah}, I. and {Paumard}, T. and {Perrin}, G. and {Zech}, A.},
        title = "{Magnetic reconnection plasmoid model for Sagittarius A* flares}",
      journal = {A\&A},
         year = 2023,
        month = apr,
       volume = {672},
          eid = {A62},
        pages = {A62},
          doi = {10.1051/0004-6361/202244936},
archivePrefix = {arXiv},
       eprint = {2301.11874},
 primaryClass = {astro-ph.HE},
       adsurl = {https://ui.adsabs.harvard.edu/abs/2023A&A...672A..62A}
}

@ARTICLE{Ripperda_2020,
       author = {{Ripperda}, Bart and {Bacchini}, Fabio and {Philippov}, Alexander A.},
        title = "{Magnetic Reconnection and Hot Spot Formation in Black Hole Accretion Disks}",
      journal = {ApJ},
         year = 2020,
        month = sep,
       volume = {900},
       number = {2},
          eid = {100},
        pages = {100},
          doi = {10.3847/1538-4357/ababab},
archivePrefix = {arXiv},
       eprint = {2003.04330},
 primaryClass = {astro-ph.HE},
       adsurl = {https://ui.adsabs.harvard.edu/abs/2020ApJ...900..100R}
}

@ARTICLE{ball:2021,
       author = {{Ball}, David and {{\"O}zel}, Feryal and {Christian}, Pierre and {Chan}, Chi-Kwan and {Psaltis}, Dimitrios},
        title = "{A Plasmoid model for the Sgr A* Flares Observed With Gravity and CHANDRA}",
      journal = {ApJ},
         year = 2021,
        month = aug,
       volume = {917},
       number = {1},
          eid = {8},
        pages = {8},
          doi = {10.3847/1538-4357/abf8ae},
archivePrefix = {arXiv},
       eprint = {2005.14251},
 primaryClass = {astro-ph.HE},
       adsurl = {https://ui.adsabs.harvard.edu/abs/2021ApJ...917....8B}
}

@article{Vos:2022,
       author = {{Vos}, J. and {Mo{\'s}cibrodzka}, M.~A. and {Wielgus}, M.},
        title = "{Polarimetric signatures of hot spots in black hole accretion flows}",
      journal = {A\&A},
         year = 2022,
        month = dec,
       volume = {668},
          eid = {A185},
        pages = {A185},
          doi = {10.1051/0004-6361/202244840},
archivePrefix = {arXiv},
       eprint = {2209.09931},
 primaryClass = {astro-ph.HE},
       adsurl = {https://ui.adsabs.harvard.edu/abs/2022A&A...668A.185V}
}

@ARTICLE{Monika2018,
       author = {{Mo{\'s}cibrodzka}, M. and {Gammie}, C.~F.},
        title = "{IPOLE - semi-analytic scheme for relativistic polarized radiative transport}",
      journal = {MNRAS},
         year = 2018,
        month = mar,
       volume = {475},
       number = {1},
        pages = {43-54},
          doi = {10.1093/mnras/stx3162},
archivePrefix = {arXiv},
       eprint = {1712.03057},
 primaryClass = {astro-ph.HE},
       adsurl = {https://ui.adsabs.harvard.edu/abs/2018MNRAS.475...43M}
}

@article{Brod-Loeb2006,
       author = {{Broderick}, Avery E. and {Loeb}, Abraham},
        title = "{Imaging optically-thin hotspots near the black hole horizon of Sgr A* at radio and near-infrared wavelengths}",
      journal = {MNRAS},
         year = 2006,
        month = apr,
       volume = {367},
       number = {3},
        pages = {905-916},
          doi = {10.1111/j.1365-2966.2006.10152.x},
archivePrefix = {arXiv},
       eprint = {astro-ph/0509237},
 primaryClass = {astro-ph},
       adsurl = {https://ui.adsabs.harvard.edu/abs/2006MNRAS.367..905B}
}

@article{W22,
        author = {{Wielgus}, M. and {Moscibrodzka}, M. and {Vos}, J. and {others}},
        title = "{Orbital motion near Sagittarius A$^{*}$ . Constraints from polarimetric ALMA observations}",
        journal = {A\&A},
        year = 2022,
        month = sep,
        volume = {665},
        eid = {L6},
        pages = {L6},
        note  = {(W22)},
        doi = {10.1051/0004-6361/202244493},
        archivePrefix = {arXiv},
       eprint = {2209.09926},
        primaryClass = {astro-ph.HE},
       adsurl = {https://ui.adsabs.harvard.edu/abs/2022A&A...665L...6W}
}

@ARTICLE{Gelles:2021,
       author = {{Gelles}, Zachary and {Himwich}, Elizabeth and {Johnson}, Michael D. and {Palumbo}, Daniel C.~M.},
        title = "{Polarized image of equatorial emission in the Kerr geometry}",
      journal = {PRD},
         year = 2021,
        month = aug,
       volume = {104},
       number = {4},
          eid = {044060},
        pages = {044060},
          doi = {10.1103/PhysRevD.104.044060},
archivePrefix = {arXiv},
       eprint = {2105.09440},
 primaryClass = {gr-qc},
       adsurl = {https://ui.adsabs.harvard.edu/abs/2021PhRvD.104d4060G}
}

@article{gravityMichi,
       author = {{GRAVITY Collaboration} and {Baub{\"o}ck}, M. and {others}},
        title = "{Modeling the orbital motion of Sgr A*'s near-infrared flares}",
      journal = {A\&A},
         year = 2020,
        month = mar,
       volume = {635},
          eid = {A143},
        pages = {A143},
          doi = {10.1051/0004-6361/201937233},
archivePrefix = {arXiv},
       eprint = {2002.08374},
 primaryClass = {astro-ph.HE},
       adsurl = {https://ui.adsabs.harvard.edu/abs/2020A&A...635A.143G}
}

@article{gravityAlejandra,
       author = {{GRAVITY Collaboration} and {Jim{\'e}nez-Rosales}, A. and {others}},
        title = "{Dynamically important magnetic fields near the event horizon of Sgr A*}",
      journal = {A\&A},
         year = 2020,
        month = nov,
       volume = {643},
          eid = {A56},
        pages = {A56},
          doi = {10.1051/0004-6361/202038283},
archivePrefix = {arXiv},
       eprint = {2009.01859},
 primaryClass = {astro-ph.HE},
       adsurl = {https://ui.adsabs.harvard.edu/abs/2020A&A...643A..56G}
}

@ARTICLE{Najafi2024,
       author = {{Najafi-Ziyazi}, Mahdi and {Davelaar}, Jordy and {Mizuno}, Yosuke and {Porth}, Oliver},
        title = "{Flares in the Galactic centre - II. Polarization signatures of flares at mm-wavelengths}",
      journal = {MNRAS},
         year = 2024,
        month = jul,
       volume = {531},
       number = {4},
        pages = {3961-3972},
          doi = {10.1093/mnras/stae1343},
archivePrefix = {arXiv},
       eprint = {2308.16740},
 primaryClass = {astro-ph.HE},
       adsurl = {https://ui.adsabs.harvard.edu/abs/2024MNRAS.531.3961N}
}

@ARTICLE{Porth2021_fluxer,
       author = {{Porth}, O. and {Mizuno}, Y. and {Younsi}, Z. and {Fromm}, C.~M.},
        title = "{Flares in the Galactic Centre - I. Orbiting flux tubes in magnetically arrested black hole accretion discs}",
      journal = {MNRAS},
         year = 2021,
        month = apr,
       volume = {502},
       number = {2},
        pages = {2023-2032},
          doi = {10.1093/mnras/stab163},
archivePrefix = {arXiv},
       eprint = {2006.03658},
 primaryClass = {astro-ph.HE},
       adsurl = {https://ui.adsabs.harvard.edu/abs/2021MNRAS.502.2023P}
}

@ARTICLE{Gravity2022,
       author = {{GRAVITY Collaboration}},
        title = "{Mass distribution in the Galactic Center based on interferometric astrometry of multiple stellar orbits}",
      journal = {A\&A},
         year = 2022,
        month = jan,
       volume = {657},
          eid = {L12},
        pages = {L12},
          doi = {10.1051/0004-6361/202142465},
       adsurl = {https://ui.adsabs.harvard.edu/abs/2022A&A...657L..12G}
}

@ARTICLE{eht:2022_paperI,
       author = {{EHT Collaboration}},
        title = "{First Sagittarius A* Event Horizon Telescope Results. I. The Shadow of the Supermassive Black Hole in the Center of the Milky Way}",
      journal = {ApJL},
         year = 2022,
        month = may,
       volume = {930},
       number = {2},
          eid = {L12},
        pages = {L12},
          doi = {10.3847/2041-8213/ac6674},
       adsurl = {https://ui.adsabs.harvard.edu/abs/2022ApJ...930L..12A}
}

@ARTICLE{eht:2022_paperII,
       author = {{EHT Collaboration}},
        title = "{First Sagittarius A* Event Horizon Telescope Results. II. EHT and Multiwavelength Observations, Data Processing, and Calibration}",
      journal = {ApJL},
         year = 2022,
        month = may,
       volume = {930},
       number = {2},
          eid = {L13},
        pages = {L13},
          doi = {10.3847/2041-8213/ac6675},
       adsurl = {https://ui.adsabs.harvard.edu/abs/2022ApJ...930L..13A}
}

@ARTICLE{eht:2022_paperV,
       author = {{EHT Collaboration}},
        title = "{First Sagittarius A* Event Horizon Telescope Results. V. Testing Astrophysical Models of the Galactic Center Black Hole}",
      journal = {ApJL},
         year = 2022,
        month = may,
       volume = {930},
       number = {2},
          eid = {L16},
        pages = {L16},
          doi = {10.3847/2041-8213/ac6672},
       adsurl = {https://ui.adsabs.harvard.edu/abs/2022ApJ...930L..16A}
}

@ARTICLE{gravity:2018,
       author = {{GRAVITY Collaboration}},
        title = "{Detection of orbital motions near the last stable circular orbit of the massive black hole SgrA*}",
      journal = {A\&A},
         year = 2018,
        month = oct,
       volume = {618},
          eid = {L10},
        pages = {L10},
          doi = {10.1051/0004-6361/201834294},
archivePrefix = {arXiv},
       eprint = {1810.12641},
 primaryClass = {astro-ph.GA},
       adsurl = {https://ui.adsabs.harvard.edu/abs/2018A&A...618L..10G}
}

@ARTICLE{Wielgus2022_LC,
       author = {{Wielgus}, Maciek and {Marchili}, Nicola and {Mart{\'\i}-Vidal}, Iv{\'a}n and {others}},
        title = "{Millimeter Light Curves of Sagittarius A* Observed during the 2017 Event Horizon Telescope Campaign}",
      journal = {ApJL},
         year = 2022,
        month = may,
       volume = {930},
       number = {2},
          eid = {L19},
        pages = {L19},
          doi = {10.3847/2041-8213/ac6428},
archivePrefix = {arXiv},
       eprint = {2207.06829},
 primaryClass = {astro-ph.HE},
       adsurl = {https://ui.adsabs.harvard.edu/abs/2022ApJ...930L..19W}
}

@ARTICLE{Ripperda_2022,
       author = {{Ripperda}, B. and {Liska}, M. and {Chatterjee}, K. and {Musoke}, G. and {others}},
        title = "{Black Hole Flares: Ejection of Accreted Magnetic Flux through 3D Plasmoid-mediated Reconnection}",
      journal = {ApJL},
         year = 2022,
        month = jan,
       volume = {924},
       number = {2},
          eid = {L32},
        pages = {L32},
          doi = {10.3847/2041-8213/ac46a1},
archivePrefix = {arXiv},
       eprint = {2109.15115},
 primaryClass = {astro-ph.HE},
       adsurl = {https://ui.adsabs.harvard.edu/abs/2022ApJ...924L..32R}
}

@Article{   dexter20,
       author = {{Dexter}, J. and {Tchekhovskoy}, A. and {Jim{\'e}nez-Rosales}, A. and {others}},
        title = "{Sgr A* near-infrared flares from reconnection events in a magnetically arrested disc}",
      journal = {MNRAS},
         year = 2020,
        month = oct,
       volume = {497},
       number = {4},
        pages = {4999-5007},
          doi = {10.1093/mnras/staa2288},
archivePrefix = {arXiv},
       eprint = {2006.03657},
 primaryClass = {astro-ph.HE},
       adsurl = {https://ui.adsabs.harvard.edu/abs/2020MNRAS.497.4999D}
}

@ARTICLE{gravity2023,
       author = {{Gravity Collaboration}},
        title = "{Polarimetry and astrometry of NIR flares as event horizon scale, dynamical probes for the mass of Sgr A*}",
      journal = {A\&A},
         year = 2023,
        month = sep,
       volume = {677},
          eid = {L10},
        pages = {L10},
          doi = {10.1051/0004-6361/202347416},
archivePrefix = {arXiv},
       eprint = {2307.11821},
 primaryClass = {astro-ph.GA},
       adsurl = {https://ui.adsabs.harvard.edu/abs/2023A&A...677L..10G}
}

@ARTICLE{Tiede2020,
       author = {{Tiede}, Paul and {Pu}, Hung-Yi and {Broderick}, Avery E. and {Gold}, Roman and {Karami}, Mansour and {Preciado-L{\'o}pez}, Jorge A.},
        title = "{Spacetime Tomography Using the Event Horizon Telescope}",
      journal = {ApJ},
         year = 2020,
        month = apr,
       volume = {892},
       number = {2},
          eid = {132},
        pages = {132},
          doi = {10.3847/1538-4357/ab744c},
archivePrefix = {arXiv},
       eprint = {2002.05735},
 primaryClass = {astro-ph.HE},
       adsurl = {https://ui.adsabs.harvard.edu/abs/2020ApJ...892..132T}
}

@ARTICLE{Wielgus2023,
       author = {{Wielgus}, Maciek and {Issaoun}, Sara and {Mart{\'\i}-Vidal}, Iv{\'a}n and {Emami}, Razieh and {Moscibrodzka}, Monika and {Brinkerink}, Christiaan D. and {Goddi}, Ciriaco and {Fomalont}, Ed},
        title = "{The internal Faraday screen of Sagittarius A*}",
      journal = {A\&A},
         year = 2024,
        month = feb,
       volume = {682},
          eid = {A97},
        pages = {A97},
          doi = {10.1051/0004-6361/202347772},
archivePrefix = {arXiv},
       eprint = {2308.11712},
 primaryClass = {astro-ph.HE},
       adsurl = {https://ui.adsabs.harvard.edu/abs/2024A&A...682A..97W}
}

@ARTICLE{Narayan2003,
       author = {{Narayan}, Ramesh and {Igumenshchev}, Igor V. and {Abramowicz}, Marek A.},
        title = "{Magnetically Arrested Disk: an Energetically Efficient Accretion Flow}",
      journal = {PASJ},
         year = 2003,
        month = dec,
       volume = {55},
        pages = {L69-L72},
          doi = {10.1093/pasj/55.6.L69},
archivePrefix = {arXiv},
       eprint = {astro-ph/0305029},
 primaryClass = {astro-ph},
       adsurl = {https://ui.adsabs.harvard.edu/abs/2003PASJ...55L..69N}
}

@ARTICLE{Vincent2023,
       author = {{Vincent}, F.~H. and {Wielgus}, M. and {Aimar}, N. and {Paumard}, T. and {Perrin}, G.},
        title = "{Polarized signatures of orbiting hot spots: Special relativity impact and probe of spacetime curvature}",
      journal = {A\&A},
         year = 2024,
        month = apr,
       volume = {684},
          eid = {A194},
        pages = {A194},
          doi = {10.1051/0004-6361/202348016},
archivePrefix = {arXiv},
       eprint = {2309.10053},
 primaryClass = {astro-ph.HE},
       adsurl = {https://ui.adsabs.harvard.edu/abs/2024A&A...684A.194V}
}

@ARTICLE{yfantis24b,
       author = {{Yfantis}, A.~I. and {Wielgus}, M. and {Mo{\'s}cibrodzka}, M.},
        title = "{Hot spots around Sagittarius A*: Joint fits to astrometry and polarimetry}",
      journal = {A\&A},
         year = 2024,
        month = nov,
       volume = {691},
          eid = {A327},
        pages = {A327},
          doi = {10.1051/0004-6361/202451884},
archivePrefix = {arXiv},
       eprint = {2408.07120},
 primaryClass = {astro-ph.HE},
       adsurl = {https://ui.adsabs.harvard.edu/abs/2024A&A...691A.327Y}
}

@ARTICLE{yfantis24a,
       author = {{Yfantis}, A.~I. and {Mo{\'s}cibrodzka}, M.~A. and {Wielgus}, M. and {others}},
        title = "{Fitting the light curves of Sagittarius A* with a hot-spot model. Bayesian modeling of QU loops in the millimeter band}",
      journal = {A\&A},
         year = 2024,
        month = may,
       volume = {685},
          eid = {A142},
        pages = {A142},
          doi = {10.1051/0004-6361/202348230},
archivePrefix = {arXiv},
       eprint = {2310.07762},
 primaryClass = {astro-ph.HE},
       adsurl = {https://ui.adsabs.harvard.edu/abs/2024A&A...685A.142Y}
}

\end{document}